\documentclass[prl,twocolumn,floatfix,showpacs,superscriptaddress]{revtex4-1}
\usepackage{graphics}
\usepackage{epsfig}
\usepackage{times}
\usepackage{bm}
\usepackage{braket}
\usepackage{color}

\usepackage{amsmath}

\begin{document}

\title{Non-Symmetry-Breaking Ground State of the $S=1$ Heisenberg Model
on the Kagome lattice}
\author{Satoshi Nishimoto}
\affiliation{Institute for Theoretical Solid State Physics, IFW Dresden,
Helmholtzstrasse 20, 01069 Dresden, Germany}
\author{Masaaki Nakamura}
\affiliation{Institute of Industrial Science, the University of Tokyo,
Meguro-ku, Tokyo, 153-8505, Japan}

\date{\today}

\begin{abstract}
The $S=1$ antiferromagnetic Heisenberg model on a Kagome lattice is
studied using the density-matrix renormalization group method. To
identify the ground state, we take four kinds of clusters into account;
periodic, cylindrical, and two open ones.  The hexagonal singlet solid
(HSS) and triangular valence bond solid (TVBS) states are artificially
generated by modulating edge shapes of the open clusters.  We find that
the energy par sites of the HSS state is $e_0=E_0/N =-1.41095$, which
is readily lower than that of the TVBS state ($e_0=-1.3912 \pm 0.0025$).
This agrees well with those of the cylindrical ($e_0=-1.40988$) and periodic
($e_0=-1.409 \pm 0.005$) clusters, where no assumption as to the
ordering is posed. Thus, we conclude that the HSS picture is consistent
to describe the ground state of the $S=1$ Kagome Heisenberg model.  This
is further confirmed by finding non-symmetry-breaking state in the
calculations of the dimer-dimer correlation functions as well as the
entanglement entropy of cylindrical clusters. Finally, we estimate the
single-triplet energy gap: The HSS ground state has $\Delta=0.1919$,
while the TVBS excited state has a larger one $\Delta=0.2797$.
\end{abstract}
\maketitle

\makeatletter

For a long time, frustrated spin systems have been fascinating subjects
of research
for discovering new physics
\cite{Moessner-R}.  Among them a system attracting the most attention in
recent years is Kagome antiferromagnetic Heisenberg (KAH) model, and a
lot of experimental and theoretical studies on the Kagome system have
been carried out, assisted by the improvement of research techniques.

In the $S=1/2$ KAH system, one of the most striking finding is the fact
that this ground state is characterized as a $Z_2$ spin liquid \cite{Yan11,Jiang12}.
The experimental realization of spin liquid has been also demonstrated in
the herbertsmithite ZnCu$_3$(OD)$_6$C$_{l2}$ \cite{Han}. Theoretically,
it is still debated whether the ground state is gapless \cite{Iqbal-B-S-P}
or gapped with very small energy gap \cite{Yan11,Jiang08,Depenbrock12,Nishimoto-S-H}.
Another notable feature is the appearance of a series of plateaus in
the magnetization process \cite{Nishimoto-S-H,Capponi13}. Of particular
interest is a possible $Z_3$ spin liquid plateau at 1/9 magnetization \cite{Nishimoto-S-H}.
Thus the $S=1/2$ KAH model exhibits a variety of phases, although
it is a simple Heisenberg system consisting only of the nearest-neighbor
exchange couplings.

\begin{figure}[t]
\centering
\includegraphics[clip,scale=0.60]{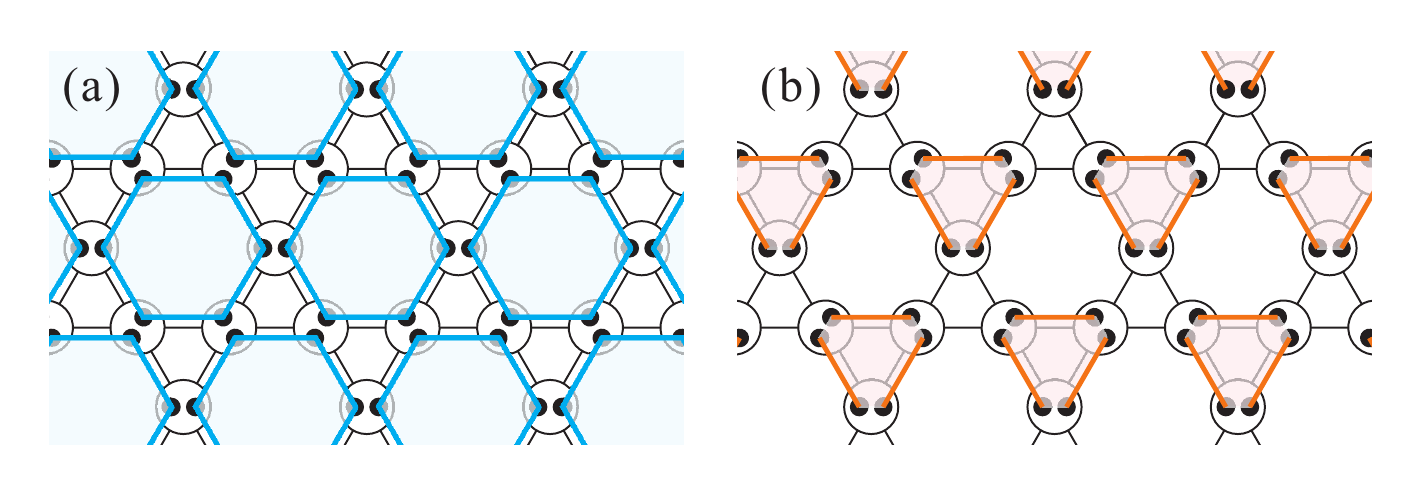}
\caption{Schematic illustrations of valence-bond solid (VBS) state in
 the $S=1$ Kagome antiferromagnetic Heisenberg model: 
 (a) the hexagonal-singlet solid (HSS) and (b) the triangular VBS states.
 Blue (red) line indicates a spin-singlet formation in six (two) $S=1/2$
 variables.}
\label{states}
\end{figure}

We here turn our attention on an $S=1$ version of the KAH model.
It would be a natural continuation of the study on the KAH system.
The Hamiltonian is written as
\begin{equation}
 H=J\sum_{\langle ij \rangle} {\bf S}_i \cdot {\bf S}_j
 \label{ham}
\end{equation}
where ${\bf S}_i$ is a spin-one operator at site $i$, and the summation
is taken for nearest neighbours $\braket{ij}$. $J$ is the
antiferromagnetic exchange integral and we set $J=1$ hereafter. From the
theoretical point of view, a very interesting point is an extension of
valence-bond solid (VBS) picture \cite{Affleck-KLT} to
two-dimension. Firstly, the triangular VBS (TVBS) state was suggested as
a ground state by the perturbation expansion around the complete TVBS
limit \cite{Asakawa}. After that, according to the analysis based on the
exact diagonalization and variational method, it was claimed that
hexagonal-singlet solid (HSS) ground state is realized \cite{Hida} (see
Fig.~\ref{states}). Moreover, a resonating AKLT loop (RAL) state has
been recently suggested as an alternative candidate~\cite{Li14}.  The
ground state of the $S=1$ KAH model is still an open issue.  So far,
several materials have been synthesized as possible realizations of the
$S=1$ KAH system, e.g., $m$-MPYNN$\cdot$BF$_4$ \cite{Wada,Matsushita},
KV$_3$Ge$_2$O$_9$ \cite{Hara},
[C$_6$N$_2$H$_8$][NH$_4$]$_2$[Ni$_3$F$_6$(SO$_4$)$_2$] \cite{Behera},
and NaV$_3$(OH)$_6$(SO$_4$)$_2$ \cite{Papoutsakis}.  Further
experimental observations are strongly expected.

\begin{figure}[hbt]
\centering \includegraphics[clip,scale=0.55]{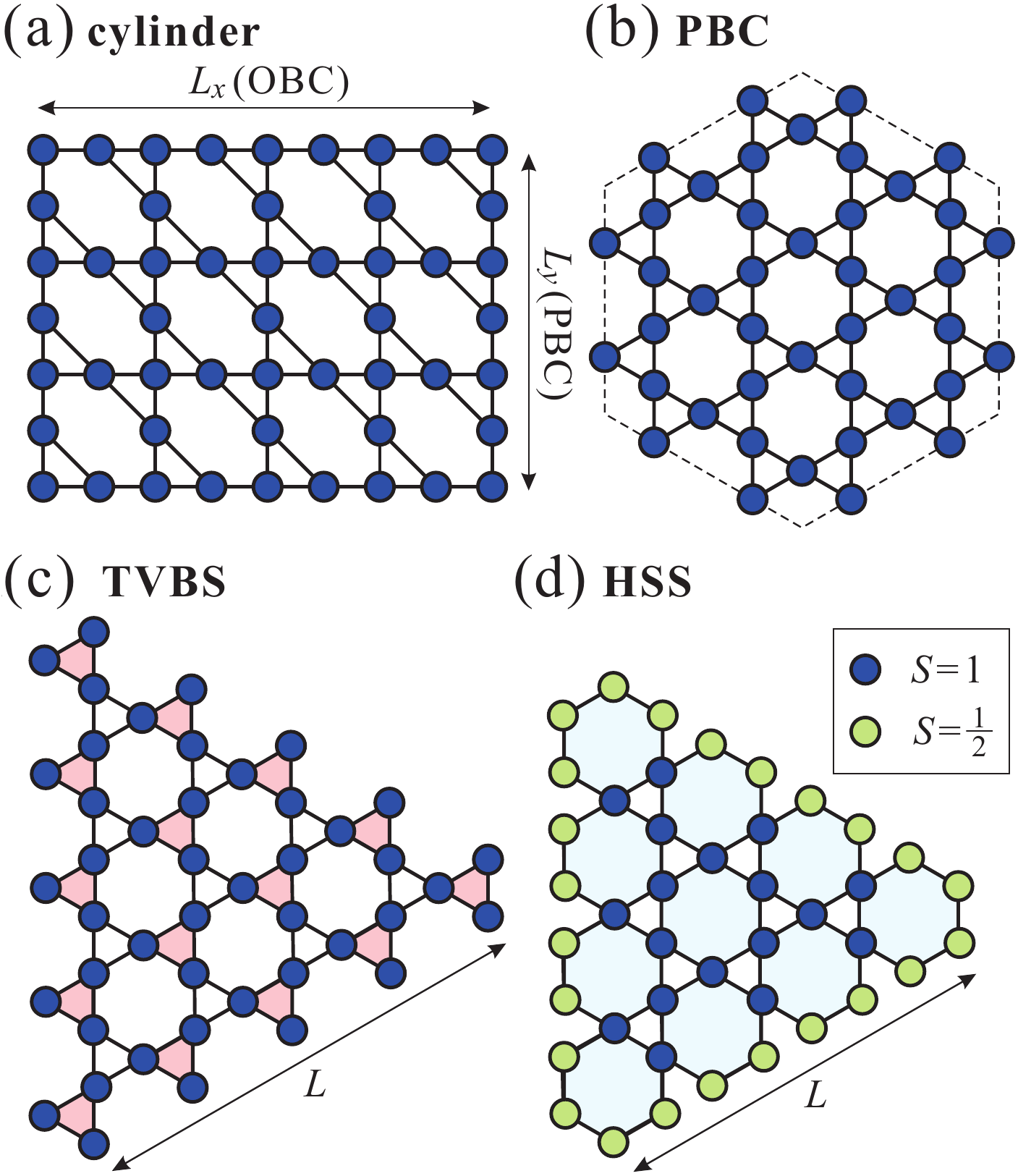}
\caption{Lattices used for the DMRG calculations:  (a) A cylindrical cluster,
denoted as XC6-3, where periodic (open) boundary conditions is applied in
the vertical (horizontal) direction. (b) A PBC cluster with $N=36$
sites. Other PBC clusters are shown in the Supplemental material.
(c),(d) Open clusters for obtaining the TVBS and HSS states, respectively.}
\label{lattice}
\end{figure}

In this Letter, the ground state of the $S=1$ KAH model is determined.
We use the density-matrix renormalization group (DMRG) method, which
enables us to study large-size clusters. For this aim, we exploit four kinds
of clusters shown in Fig.~\ref{lattice}. Up to about 10000 density-matrix
eigenstates are kept in the renormalization procedure and the discarded
weight is below 10$^{-4}$ even for the most difficult case, namely,
36-site cluster under the periodic boundary conditions (PBC). First, by
calculating the dimer-dimer correlation functions and entanglement entropy
with the cylindrical clusters, we find that the translational symmetry is not
broken in the ground state.It is further confirmed with the PBC cluster.
Next, we {\it intentionally} produce the HSS and TVBS states by modulating
the edge condition of open clusters (see below for details). It enables us
to compare their energies directly. The lowest-state energy calculated
with the HSS cluster is $E_0/N \equiv e_0 =-1.41095$ and it is in good
agreement with those of the cylindrical ($e_0=-1.40988$) and PBC
($e_0=-1.409 \pm 0.005$) clusters. It is quite reasonable because the
HSS state is a non-symmetry-breaking one. Moreover, it is striking that
the energy estimated with the TVBS cluster ($e_0=-1.3912 \pm 0.0025$) is
decidedly higher than the others. Thus, we verify the HSS state to be the
ground state of the $S=1$ KAH model. Finally, we estimate the single-triplet
gap. The ground state, i.e., the HSS state, has a gap $\Delta=0.1919$;
while, the TVBS state as an excited eigenstate has a larger gap $\Delta=0.2797$.

\begin{figure*}[tb]
\centering \includegraphics[clip,scale=0.85]{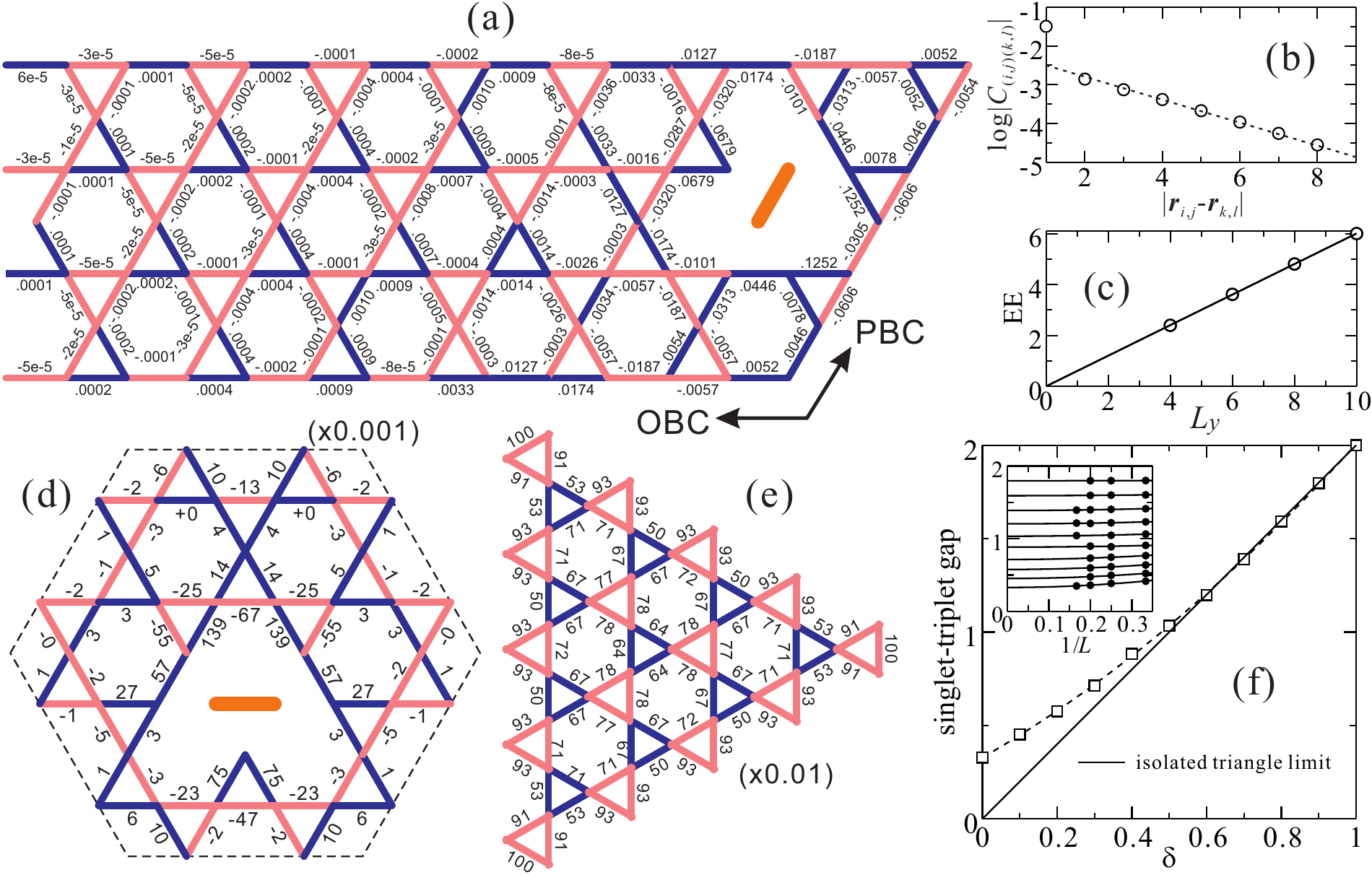} \caption{(a)
Dimer-dimer correlation functions for the XC6-3 cylinder with $L_y=24$. The
reference bond is indicated by the thick orange line. (b) Semilog plot
of the absolute value of dimer-dimer correlation functions with distance
from the reference bond.  (c) Entanglement entropy with $L_y$. The $L_x
\to \infty$ limit has been already taken.  (d) Dimer-dimer correlation
functions for the PBC cluster ($N=36$).  (e) Nearest-neighbour spin-spin
correlation functions for the TVBS cluster.  (f) Singlet-triplet
gap extrapolated to the thermodynamic limit for the modulated TVBS
clusters (see text).  Inset: the extrapolation scheme as a function of
$1/L$.  } \label{dimerdimer}
\end{figure*}
 
Let us start with the cylindrical cluster [Fig.~\ref{lattice}(a)].
We here use a type of cylinder denoted as XC6-3, the notation of which
was defined in previous works on the $S=1/2$ KAH system~\cite{Yan11,Depenbrock12}.
The reason for choosing it is related to the shape of both edges.
In general, an open edge exerts a critical influence on the formation
of plaquette or bond singlets. If the edge consists of either triangles
or hexagons, the TVBS or HSS state may be artificially favoured in our
system (\ref{ham}). Such a signature was also identified in the $S=1/2$
KAH model~\cite{Yan11,Gong13}. This problem should be avoided by choosing
a XC$n$-($n/2$) or YC$n$ type of cylinder, where both triangles and hexagons
are {\it equally} arranged at the open edges (see Supplement). In other words,
the TVBS and HSS states could be intentionally stabilized by modulating
the shape of open edges in a (small) cluster. This {\it technique} is used
in the latter part of this Letter. For instance, the same technique was
used to detect the plaquette VBS state in the $J_1$-$J_2$ honeycomb
Heisenberg model~\cite{Ganesh13}.

In order to check the configuration of singlet valence bonds, we calculate
the dimer-dimer correlation functions defined by
\begin{equation}
 C_{(i,j)(k,l)}=
  4[\braket{({\bf S}_i \cdot {\bf S}_j)({\bf S}_k \cdot {\bf S}_l)}
  -\braket{{\bf S}_i \cdot {\bf S}_j}\braket{{\bf S}_k \cdot {\bf S}_l}].
\end{equation}
In Fig.~\ref{dimerdimer}(a) the values of the dimer-dimer correlation
for each bond are shown with fixing one reference dimer bond.
Blue and red links denote positive and negative correlations.
It seems that the patterns for the sign of correlations does not
exhibit any spatial periodicity. In addition, the correlation decays exponentially
with distance of two bonds, as plotted in Fig.~\ref{dimerdimer}(b).
It clearly indicates the absence of symmetry-breaking order associated
with dimer formations. A similar feature is also observed in the PBC clusters
[Fig.~\ref{dimerdimer}(d)] Moreover, to make sure, we examine
the von Neumann entanglement entropy (EE)
$S_L(l)=-{\rm Tr}_l \rho_l \log \rho_l$, where $\rho_l={\rm Tr}_{L-l}\rho$
is the reduced density matrix of the subsystem and $\rho$ is the full density
matrix of the whole system~\cite{Oshikawa06,Jiang12}. We plot the value
of $S_{\frac{L}{2}}$ as a function of the circumference of the cylinder
$L_y$. The values should follow a relation $S(L_y)=\alpha L_y - \gamma$, 
where $\alpha$ is a constant and $\gamma = \ln D$ is the topological entropy
with dimension $D$~\cite{Kitaev06,Levin06}. By the fitting of our data with
the equation, we obtain $\gamma=0.0014$ in the $L_y\to 0$ limit. This
suggests that the system is in a topologically trivial phase, i.e., a
unique ground state, and it is consistent to the results of the
dimer-dimer correlation functions.

Next, we consider the TVBS state. As mentioned above, an ordered state
like the symmetry-breaking TVBS state can be forcibly stabilized as the
lowest state in a small cluster by taking a proper edge condition.
However, we have to take notice the following two points to determine if
the lowest state is really the ground state when this artificial
technique is applied: (i) the ordering survives in the thermodynamic
limit, (ii) the energy of the ordered state remains lowest among all
eigenstates in the thermodynamic limit. If it is not the case, a level
crossing with the true ground state occurs at some larger cluster.  One
possible realization of the TVBS cluster is shown in
Fig.~\ref{lattice}(c).  For this cluster we present the values of the
nearest-neighbour spin-spin correlations $\braket{-{\bf S}_i \cdot {\bf
S}_j}$ in Fig.~\ref{dimerdimer}(e).  A link with larger (smaller) value
than that of the next bonds is coloured in red (blue). A TVBS
configuration is obviously seen. Furthermore, with increasing the system
size, the spin gap is smoothly scaled to a finite value $\Delta=0.2797$
at $1/L\to 0$ (see the inset of Fig.~\ref{dimerdimer}(f)).  The spin gap
is evaluated as the energy difference between the lowest singlet and the
first excited triplet states. Hence, the TVBS state is confirmed to
survive in the thermodynamic limit. We also make certain that this state
is indeed of the TVBS with triangular three-dimer formations.  For this
purpose, we modulate the exchange couplings as $(1+\delta)J$ and
$(1-\delta)J$ for red and blue bonds in Fig.~\ref{dimerdimer}(e),
respectively ($0 \le \delta \le 1$). Figure~\ref{dimerdimer}(f) shows
the spin gap as a function of $\delta$. In the isolated triangle limit
($\delta=1$), the lowest state is exactly described by a direct product
of triangular Haldane-gap dimers with three $S=1$ spins; as $\delta$
decreases, the fluctuations between the triangles are increased and the
original Hamiltonian is recovered at $\delta=0$. It is worth noting the
spin gap is scaled perfectly by that assuming an isolated triangle,
$\Delta=2\delta$, for a wide range of $\delta(>0.5)$. This means that
the triangular Haldane-gap dimers are quite robust against the
inter-triangle fluctuations and the product state may be a good
approximation of the TVBS state even at smaller $\delta$. We can further
see that no gap closing point exists from $\delta=1$ to $\delta=0$,
suggesting that the product state is intermittently connected to the
TVBS state in the original model (\ref{ham}).  Therefore, we can confirm
that the TVBS state indeed exists as an eigenstate of the $S=1$ KAH
model.

\begin{figure}[tbh]
\centering
\includegraphics[clip,scale=0.50]{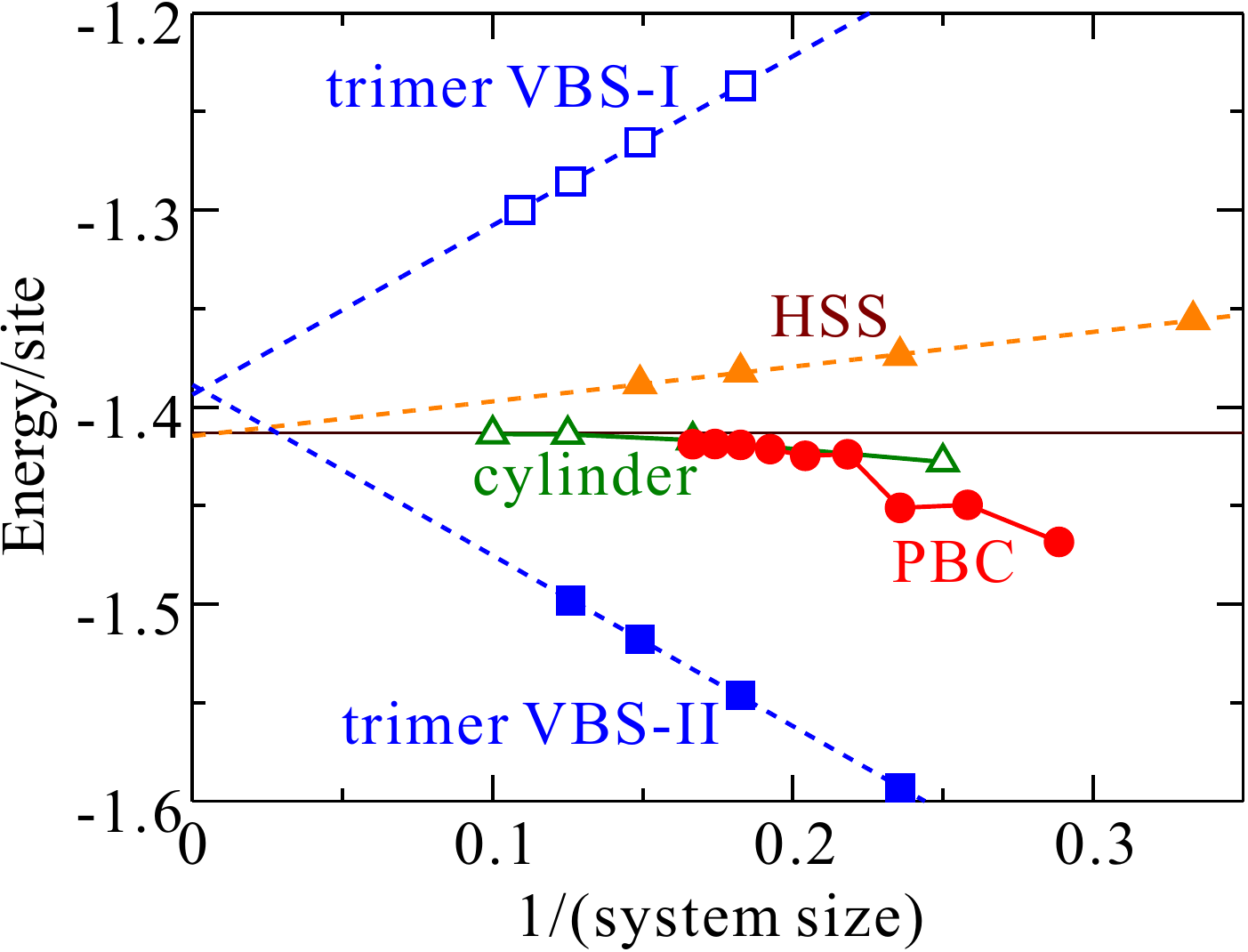}
\caption{Extrapolation scheme of the lowest-lying-state energy as a
function of $1/L_y$ for the cylindrical cluster and of $1/\sqrt{N}$ for
the TVBS, HSS, and PBC clusters.  For the cylindrical cluster the
$L_x \to \infty$ limits are already taken.}
\label{gsenergy}
\end{figure}

For identifying the true ground state, it would be a natural step to
compare the lowest-state energies calculated with different clusters.
In Fig.~\ref{gsenergy} the finite-size scaling analysis of the energy per site
for each cluster is presented.  For the TVBS and PBC clusters, the energy is
plotted with $1/\sqrt{N}$ as usually assumed for two-dimensional systems.
For the cylindrical cluster, we first take the $1/L_x\to 0$ limit followed
by the $1/L_y\to 0$ limit. Only the scaling with $1/L_y$ is shown in Fig.~\ref{gsenergy}.
We estimate the energy for the TVBS cluster in two different ways;
one is simply the total energy divided by the total number of sites
($e_0(N)=E_0(N)/N$) and the other is an average of the neighbouring
spin-spin correlations ($e_0(N)=2\overline{\langle {\bf S}_i {\bf S}_j \rangle}$,
where the factor 2 comes from the ratio of the number of sites and bonds
in the thermodynamic limit). Since one of them is extrapolated from the
higher energy side with decreasing $1/L_y$ and the other from the lower side,
as seen in Fig.~\ref{gsenergy}, this should make the scaling analysis more reliable.
The extrapolated values from the both ways agree very well in the thermodynamic
limit and it is estimated as $e_0=-1.3912 \pm 0.0025$. The energies for the
cylindrical and PBC clusters seem to converge rather faster with the system
sizes, and they are $e_0=-1.40988$ and $e_0=-1.409 \pm 0.005$,
respectively, in the thermodynamic limit. Clearly, the energy of the TVBS state
is high in number ($\Delta e_0 =0.02$) compared to those for the other two
clusters. We thus argue that the TVBS state exists as an eigenstate but
it is not the ground state.

In a similar way as the TVBS state, we can also stabilize the HSS state
in a small cluster. One possible realization is shown in Fig.~\ref{lattice}(d),
where the hexagons are placed at the corners of open cluster. Note that
the outer $S=1$ spins are replaced by $S=1/2$ spins not to hold extra free
spins when all hexagons form singlet plaquettes. By doing this we can easily 
detect the first excited triplet state by one spin-flip from the lowest
state. For the HSS cluster we estimate the energy as an averaged value of 
the neighbouring spin-spin correlations between the $S=1$ spins, namely,
the outer bonds are excluded. The size-scaling is shown in Fig.~\ref{gsenergy}.
The extrapolated value to the thermodynamic limit is $e_0 =-1.41095$.
It agrees to those of the cylindrical and PBC clusters within the error bars.
Since the HSS state is a non-symmetry-breaking one, it is utterly reasonable. 

Although it is not possible to study a RAL state with our method, our estimation
of the energy of the HSS state is readily lower than the variational energy
of the pure RAL state $e_0=-1.383$ \cite{Li14}. Therefore, we conclude that
the HSS state is the ground state of the $S=1$ KAH model.
It has been suggested that the TVBS state is the ground state 
when a certain amount of the second- and third-neighbour antiferromagnetic
exchange interactions are taken into account~\cite{Cai09}. It is likely to
happen because a frustration induced by the second-neighbour interactions 
naively lifts up the energy of the HSS state.

\begin{figure}[t]
\centering
\includegraphics[clip,scale=0.50]{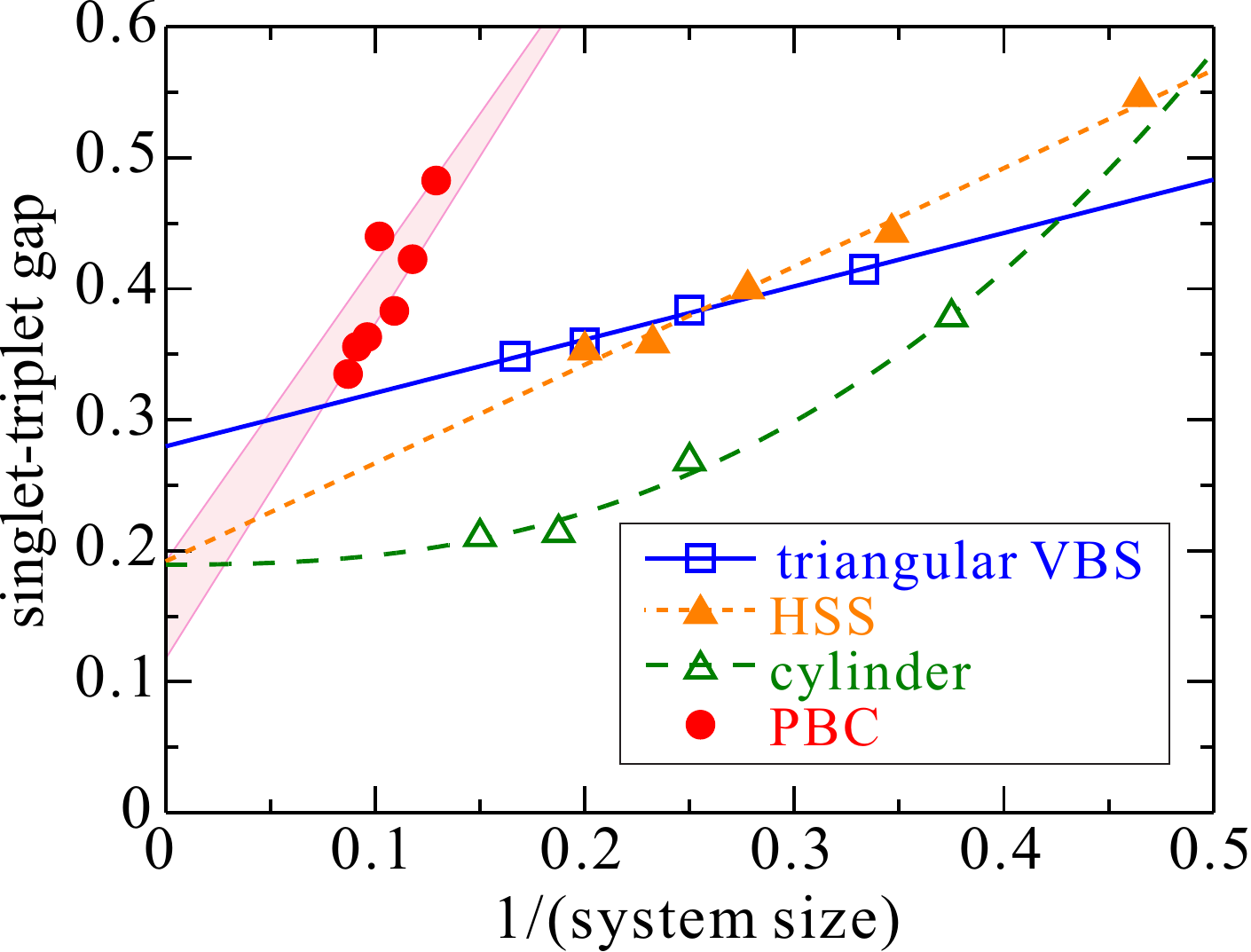}
\caption{Extrapolation scheme of the spin gap as a function
of $1/L_y$ for the cylindrical cluster, of $1/L$ for the TVBS, 
HSS, and of $1/\sqrt{N}$ for the PBC clusters. 
For the cylindrical cluster the limit for $L_x \to \infty$ are 
already taken.
}
\label{gap}
\end{figure}

Finally, we study the spin gap. In Fig.~\ref{gap} the finite-size-scaling
analyses of the spin gap for the used four kinds of clusters are performed.
The extrapolated values to the thermodynamic limit are $\Delta=0.183$,
$0.17 \pm 0.03$, and $0.172$ for the cylindrical, PBC, and HSS clusters,
respectively. As expected from the above discussion, the three numbers
are very close to each other. We thus determine the spin gap of the HSS
state, which is the ground state, is $\Delta=0.178 \pm 0.05$.  It is considerably
smaller than that of the TVBS state, i.e., $\Delta=0.2797$. This may
imply that the triangular Haldane-gap dimers in the TVBS state is more
robust than the hexagonal singlet in the HSS state. Hence, in the TVBS
state the spins are strongly screened and the energy gain from the
quantum fluctuations between the triangles might be small. Here we
shall comment on the spin gap observed in a related material
$m$-MPYNN$\cdot$BF$_4$.  From the fitting of measured susceptibility
with $(\Delta/k_B T)\exp(-\Delta/k_B T)$ at low temperature $T$,
$\Delta$ was obtained as $\sim 0.2$K~\cite{Wada,Matsushita}.  This
system can be mapped to the $S=1$ KAH model with $J=0.65-0.95$ K.
By our analysis the spin gap is estimated as $\Delta=0.178J \approx
0.12-0.17$K. This value seems to be reasonably close to the experimental one.

In summary, we have studied the $S=1$ KAH model using the DMRG
technique.  Four kinds of clusters have been used to determine the
ground state.  We have succeeded in extracting the ordered HSS and TVBS
states by taking unique open clusters. It enabled us to extrapolate
their low-lying energies individually to the thermodynamic limit. As a
result, we have found that the ground state of the $S=1$ KAH model is
the HSS state and the TVBS state is an excited one. However, their
lowest-lying-state energies are very close, and the near degeneracy
seems to make it more difficult to detect the true ground state. The
dimer-dimer correlation functions and the entanglement entropy for the
cylindrical and PBC clusters, where no assumptions for ordering are posed,
suggest a non-symmetry-breaking ground state. It also supports the HSS
ground state. The singlet-triplet gap of the TVBS state is larger than
that of the HSS state. It means that the triangular singlet dimers are
more robust than the hexagonal singlet, and the spin are strongly
screened. It may prevent lowering of the energy derived by quantum
fluctuations between triangular singlet dimers.

We acknowledge Max-Planck-Institut f\"{u}r Festk\"{o}rperforschung
where a part of the numerical calculations has been done.

{\it Note added} --- In preparing this manuscript we noticed two
preprints on the DMRG study of the $S=1$ KAH
model~\cite{Changlani14,Liu14}.  Although they argued different ground
state from our conclusion, the ground-state energy in their calculations
agrees with ours very well.

\section{cylindrical clusters}

In this Letter, we follow the notation of cylindrical cluster in 
Ref.~2. A cylinder is labeled as XC$n$ (YC$n$) when the circumference
is along or close in orientation to the $x$($y$)-axis, where $n$ is
the circumference in unit of the lattice spacing. If the lattice is
connected with a shift $d$ units in the periodic direction, the number
of shift is added to the label such as XC$n$-$d$.

\begin{figure}[h]
\center
\includegraphics[scale=0.45]{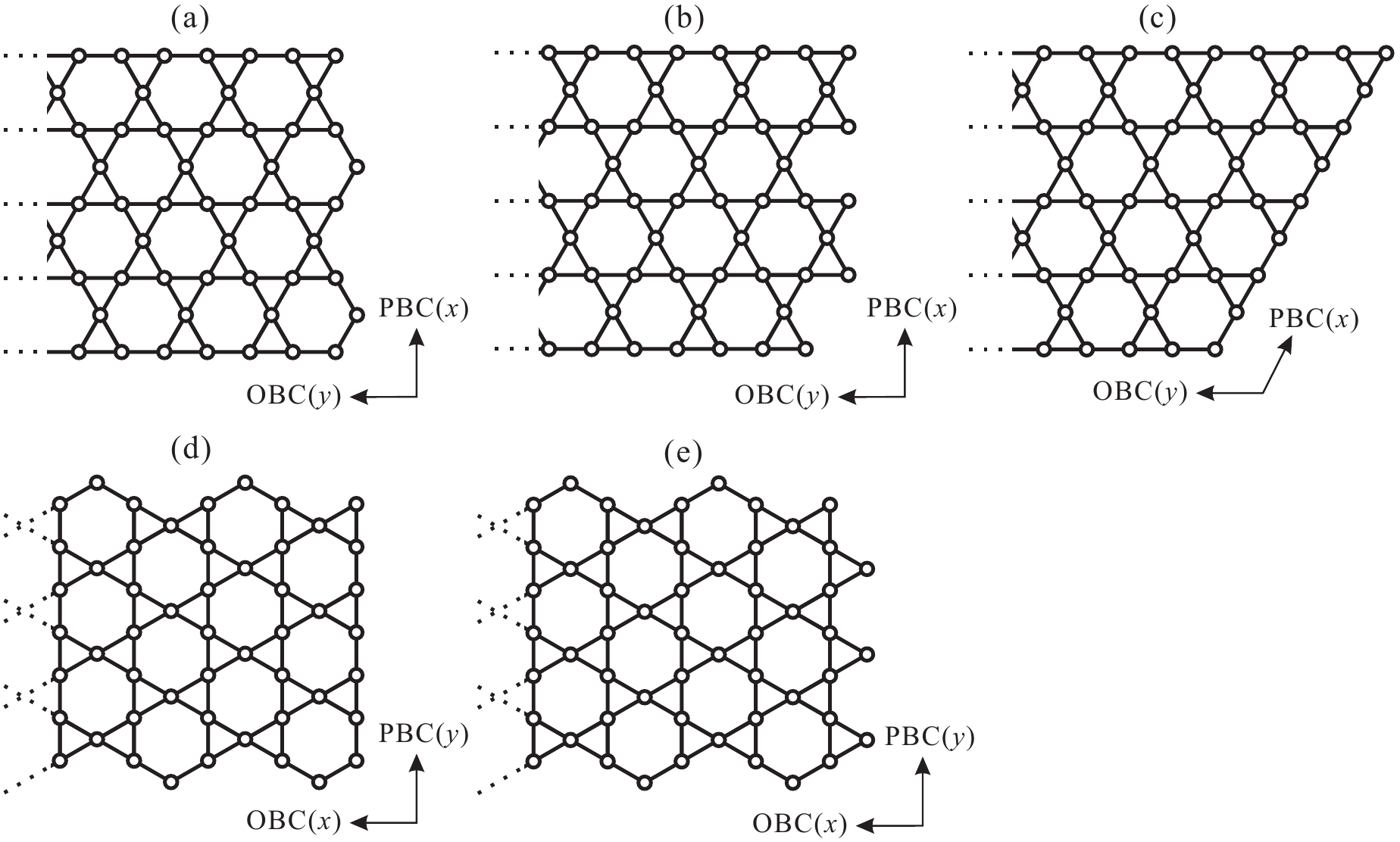}
\caption{(a),(b) Kinds of XC8 cylinders. (c) A kind of XC8-4 cylinder. (d),(e) Kinds of YC6 cylinders.
}
\label{suppl_fig1}
\end{figure}

In Fig.~\ref{suppl_fig1} several kinds of the cylindrical cluster are
illustrated. Some arbitrariness remains on the shape of open edges.
For example, the lattices (a) and (b) are both labeled by the XC8
cylinders but the shapes of the open edges are different. One of
them has hexagons and the other has triangles at the open edges.
Then, there is a possibility that the hexagonal singlet or triangular
Haldane-gap dimers is artificially favoured. If we choose the XC$n$-($n/2$)
type of cylinder, the hexagons and triangulars may be {\it equally}
placed at the open edges, as shown in Fig.~\ref{suppl_fig1}(c).
Also, a kind of the YC$n$ cylinder [Fig.~\ref{suppl_fig1}(d)] has
the similar edges. Another kind of the YC$n$ cylinder like in
Fig.~\ref{suppl_fig1}(e) has triangles at the open edges, and however,
it could be useful cluster to detect the TVBS state directly as
a translation-symmetry-breaking state along the OBC direction.

\section{periodic clusters}

\begin{figure}[h]
\center
\includegraphics[scale=0.5]{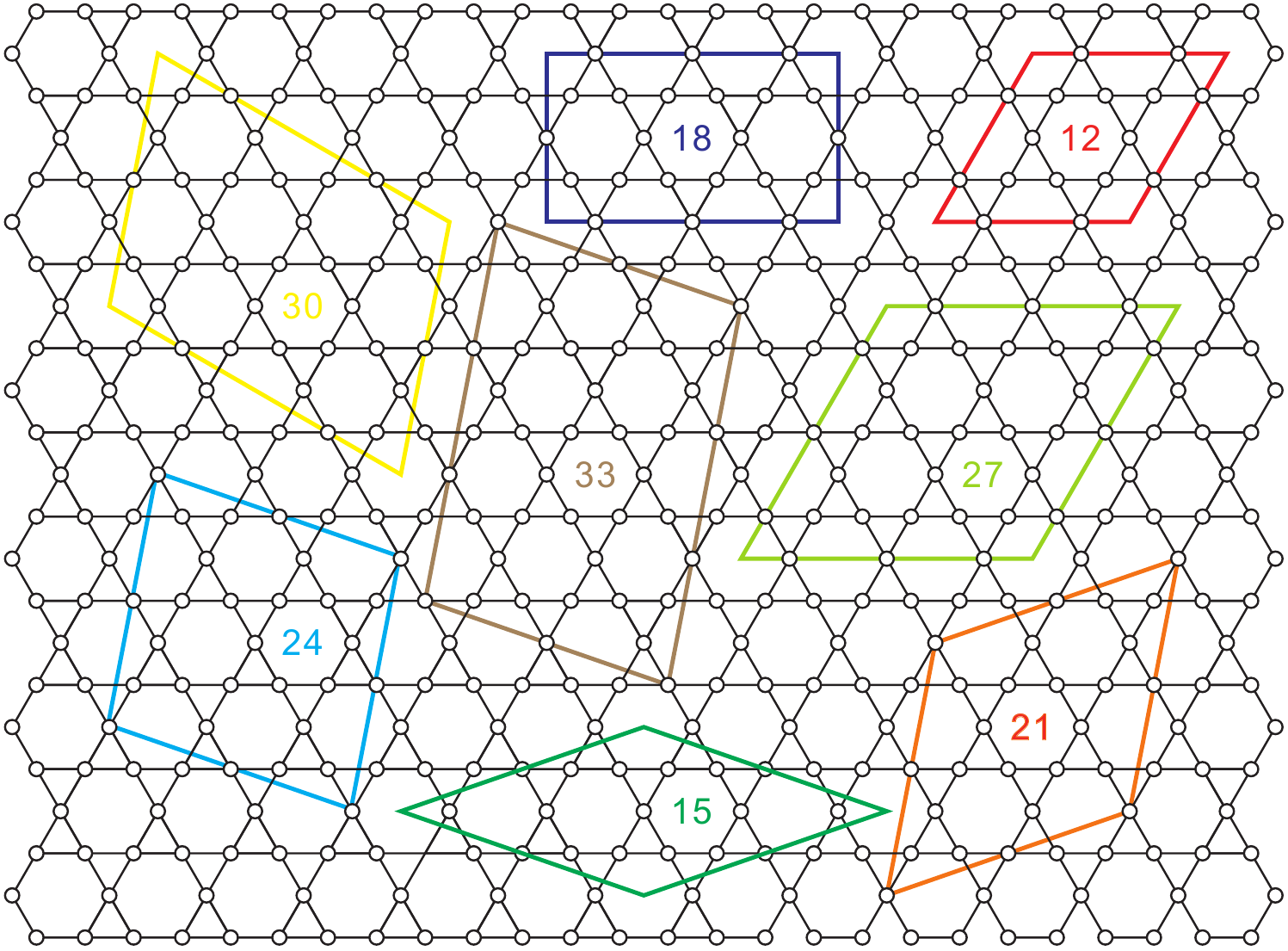}
\caption{Periodic clusters used in our numerical calculations.
}
\label{suppl_fig2}
\end{figure}

In Fig.~\ref{suppl_fig2} the periodic clusters used in our numerical calculations
are shown. They are taken as isotropic as possible. As shown in the main text, 
the $N=36$ periodic cluster is completely isotropic.

\end{document}